\documentclass[prx,aps,amsmath,amssymb,superscriptaddress]{revtex4-1}
\usepackage[english]{babel}
\usepackage{graphicx}
\newcommand{\ketbra}[1]{|#1\rangle\langle #1|}

\begin{document}

\title{Contribution of electron-phonon coupling to the luminescence spectra of single colloidal quantum dots }
\author{Eduard~A.~Podshivaylov}
\affiliation{Lomonosov Moscow State University, 119991 Moscow, Russia}
\author{Maria~A.~Kniazeva}
\affiliation{Lomonosov Moscow State University, 119991 Moscow, Russia}
\author{Aleksei~A.~Gorshelev}
\affiliation{Institute of Spectroscopy RAS, 108840 Moscow, Russia}
\author{Ivan~Yu.~Eremchev}
\email{eremchev@isan.troitsk.ru}
\affiliation{Institute of Spectroscopy RAS, 108840 Moscow, Russia}
\author {Andrei~V.~Naumov}
\affiliation{Institute of Spectroscopy RAS, 108840 Moscow, Russia}
\affiliation{Moscow State Pedagogical University, 119991 Moscow, Russia}
\author{Pavel~A.~Frantsuzov}
\email{pavel.frantsuzov@gmail.com}
\affiliation{Lomonosov Moscow State University, 119991 Moscow, Russia}
\affiliation{Voevodsky Institute of Chemical Kinetics and Combustion SB RAS, 630090 Novosibirsk, Russia}

\begin{abstract}
Luminescence spectroscopy experiments were realized for single colloidal quantum dots CdSe/ZnS in a broad temperature range above room temperature in a nitrogen atmosphere. Broadening and shifts of spectra due to the temperature change as well as due to spectral diffusion processes were detected and analyzed. A linear correlation between the positions of maxima and the squared linewidths of the spectra was found. This dependence was explained by a model which takes into account the slow variation of the electron-phonon coupling strength.
\end{abstract}
\date{\today}
\maketitle

\section{Introduction}

Colloidal semiconductor quantum dots (QDs) are very interesting objects because of their unique optical properties such as a wide absorption spectrum, a narrow emission line, a size-tunable emission wavelength, high photostability and high fluorescence quantum yield. The very first spectroscopic measurements of single CdSe quantum dots photo- luminescence revealed interesting phenomena such as long-term fluctuations of the emission intensity (blinking) \cite{BrusNature1996}  and very slow spectral diffusion (SD) \cite{BawendiPRL1996,SionnestAPL1996,BawendiPRL2013} with characteristic time scales of up to hundreds of seconds. It was shown that at cryogenic temperatures the observed emission spectrum linewidth of single QDs depends on the signal accumulation time due to spectral shifts
\cite{BawendiPRL1996,BawendiJPCB1999}, while the linewidths and the peak positions of the spectra are correlated \cite{BawendiScience1997}.

Spectral diffusion at higher temperatures was observed by Muller et al. \cite{WellerPRL2004,WellerPRB2005} in single CdSe QDs capped by a CdS rod-like shell. The linewidth and the peak position of the emission spectrum is found to be correlated at 5 K, 50 K and room temperature. These correlations at all temperatures were explained \cite{WellerPRL2004,WellerPRB2005} by the motion of the net surface charge which induces a Stark shift of the emission energy depending on the distance to the CdSe core while the spatial jitter of the charge density causes spectral line broadening. Gomez et al.\cite{MulvaneyAPL2006} noted that this hypothesis does not apply to the spherically symmetric QDs. Besides it should lead to variations of the linewidth with a change in the dielectric properties of the medium. A series of spectroscopic experiments were performed on single spherical QDs spin-coated on top of thin films of various polymer matrices at room temperature. It was shown \cite{MulvaneyAPL2006}  that there is a correlation between the linewidth and the peak position of the emission spectrum in these particles without a significant dependence on the dielectric permittivity of the matrix. Based on this, it was concluded in Ref.\cite{MulvaneyAPL2006} that the mechanism responsible for the correlated broadening and the peak position shift of the emission spectra in the PL has to be intrinsic to the QD core.

Note that the broadenings of a single QD emission spectra at 5 K and at room temperature are different in nature. At 5 K the zero-phonon line is observed and its width is much smaller than the longitudinal optical (LO) phonon energy \cite{DzhaganJPD2018}. The linewidth at room temperature becomes greater than the energy of the LO phonons, which means that the multi-phonon nature of the broadening should be taken into account \cite{HuangPRSL1950,KuboPTP1955}. While the electron-phonon coupling and spectral diffusion contributions to the spectra of chromophore molecules in solid matrices have been studied in detail
\cite{NaumovUP2018,EremchevJCP2009,KarimulinJL2014}, the same contributions in QDs are still of much interest. Here we present an in-depth experimental and theoretical study of the discussed spectral characteristics of single colloidal semiconductor QDs CdSe/ZnS in the context of their feasible relation to electron-phonon coupling.

\section{Experiment}

We performed a set of spectroscopic experiments with single QDs, including measurements with slow heating and cooling of a sample.

\begin{figure}
 \centering
 \includegraphics[width=\linewidth]{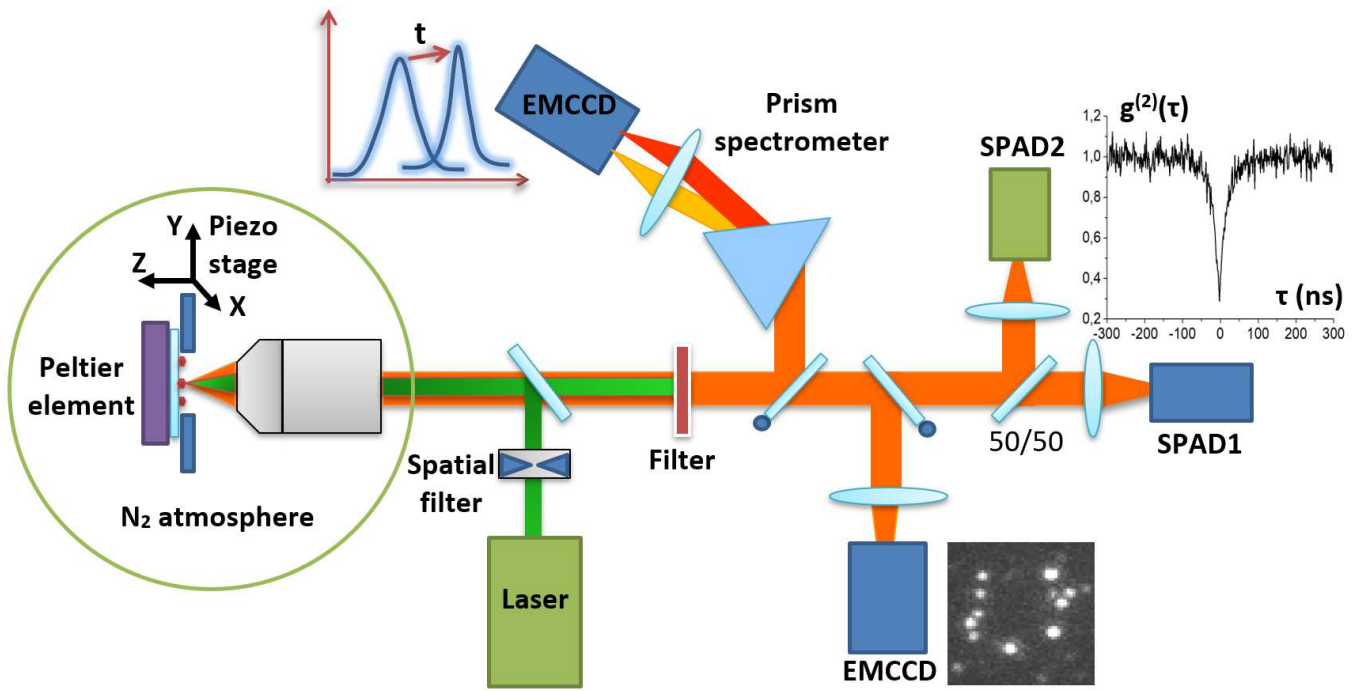}
 \caption{A schematic picture of the experimental setup}
\end{figure}

Fluorescence images and spectra of single quantum dots were recorded using a home-built fluorescence microscope equipped with a prism spectrometer \cite{NaumovJL2018,NaumovUP2018}. Two optical schemes - a wide-field scheme and a scanning confocal one (see Fig. 1) - were combined in the microscope in order to simplify the procedure of single quantum dot preliminary searching (by using fluorescence image processing and antibunching identification) and to perform sequential measurements of the fluorescence spectra of the selected QD.
Quantum dots (CdSe/ZnS from Sigma Aldrich with the fluorescence peak at 620 nm) were dispersed in a toluene solution of polyisobutylene of low concentration and then spincoated onto a cover glass. The thickness of the polymer films with single quantum dots varied within the range of several tens of nanometers. The sample was placed onto the piezo-driven stage (NanoScanTechnology), which allowed one to move the selected QD to the laser spot position with high (nanometer) precision. Between the sample and the piezo-driven stage a thermo-insulating (fluoroplastic) substrate a few millimeters thick was placed, with a hole in the center allowing the microscope objective to approach the plane of the sample at the required distance. The thermo-insulating substrate contained a temperature sensor that had good thermal contact with the sample. On top of the sample, a three-stage thermoelectric module was pressed, which was used to heat or cool the sample. This optical scheme (including the piezo-driven stage with the sample, the microscope objective, and the thermoelectric module) was mounted inside a special home-built chamber, allowing measurements both in a vacuum or in a gas nitrogen/helium atmosphere. In this particular case, the measurements were performed in a nitrogen atmosphere. The sample temperature was controlled by a LakeShore temperature controller. A tunable dye laser (Coherent CR599) or solid state laser Coherent Verdi  were used to excite quantum dots at the wavelength of 580 nm (near the quantum dot absorption band edge) or at 532 nm correspondingly.  The excitation laser intensity ($\sim$ 100 W/cm$^2$ in a focused spot) was attenuated by neutral spectral density filters (Standa) and controlled by a Newport power meter. A set of interference filters (Semrock and Thorlabs) was used for the separation of the QD fluorescence signal from the scattered laser radiation. Two highly sensitive cooled  electron multiplying charge-coupled device (EMCCD) cameras were utilized to record single quantum dot images (Andor Luca) and spectra (Andor Ixon Ultra). The Hanbury Brown and Twiss scheme with broadband 50 \% splitter (Thorlabs) and two identical  single-photon avalanche diode (SPAD) detectors (EG\&G SPCM-200PQ, time resolution 1.3 ns, dead time 200 ns, QE  65 \%) was used to measure the autocorrelation function for QD fluorescence intensity. Each fluorescence spectrum from a single quantum dot was measured with an exposure time of 200 ms and a spectral resolution of ~ 0.7 nm, which was sufficient to achieve a good signal-to-noise ratio.

\section{Results}

In the experiments at room temperature for each studied single QD we registered 2500- 3000 emission spectra with 200 ms accumulation time.
The presence of both blinking and spectral diffusion processes can be clearly seen.
Spectral traces for two QDs are shown in Fig. 2.

\begin{figure}
\centering

\begin{minipage}[h]{\linewidth}\flushleft{(a)}
\center{\includegraphics[width=1\linewidth]{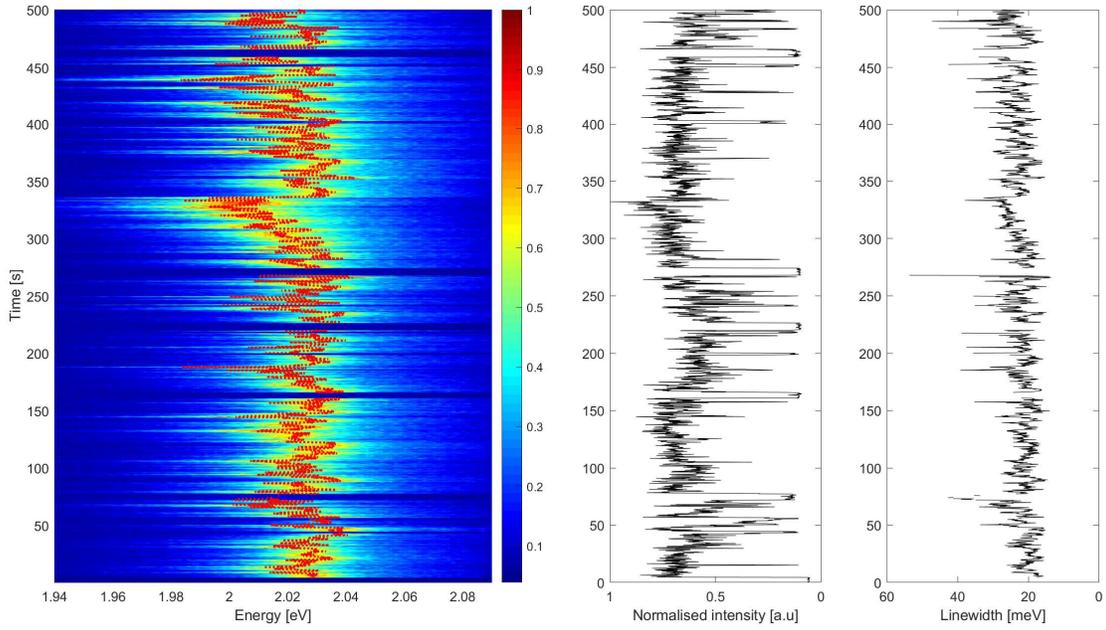}}
\end{minipage}
\hfill
\begin{minipage}[h]{\linewidth}\flushleft{(b)}
\center{\includegraphics[width=1\linewidth]{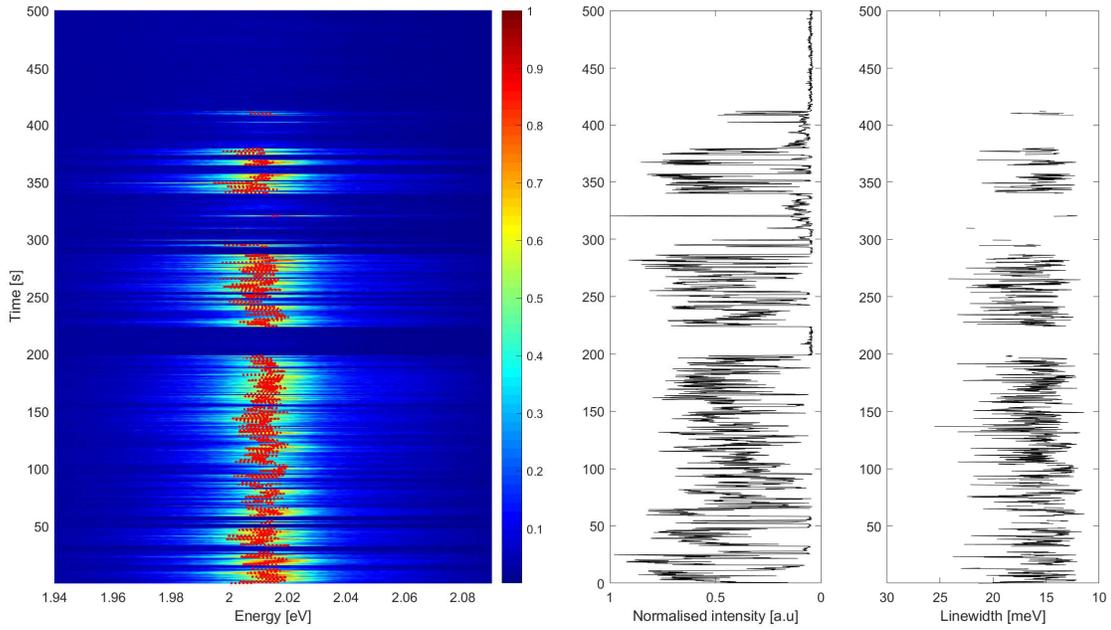}}
\end{minipage}
 \caption{ Spectral traces (left panel) where the peak position is shown by read line, time dependencies  of the normalized PL intensity (central panel) and the linewidth (right panel)  for two different single CdSe/ZnS quantum dots (a) and (b) measured at room temperature  }
\end{figure}

Each spectrum was fitted with a Gaussian function
 $$ G(\epsilon)=\frac{G_0}{\sqrt{2\pi}\sigma}\exp\left\{-\frac{(\epsilon-\epsilon_0)^2}{2\sigma^2}\right\}+b $$
 whose four parameters were: peak emission photon energy $\epsilon_0$, linewidth $\sigma$, amplitude $G_0$ and background level $b$. An example of a typical spectrum fitting is shown in Fig. 3.
\begin{figure}
 \centering
 \includegraphics[width=0.9\linewidth]{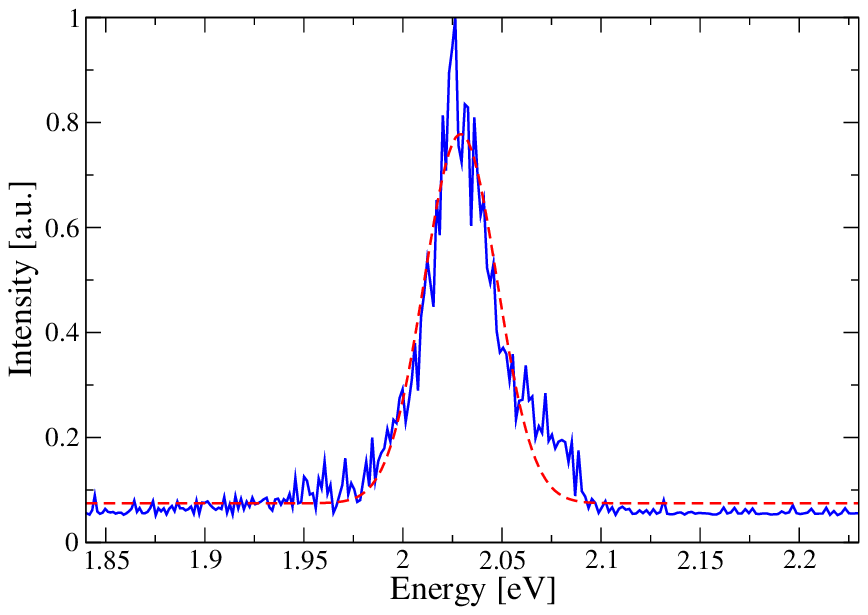}
 \caption{ The emission spectrum of a single CdSe/ZnS QD (blue line) and its fit with a Gaussian (dashed line). The parameters of the fit
 are $\epsilon_0=2.029$ eV, $\sigma=18.3$ meV }
\end{figure}

The correlation between the peak energy and the linewidth was found for all studied QDs.
As seen on Fig. 4 the peak energy dependence of the linewidth squared
can be fitted by the linear function
\begin{equation}
\sigma^2=\alpha kT (E_0-\epsilon_{0})
 \label{depend}
\end{equation}
where $T$ is the absolute temperature, $k$ is the Bolzmann constant,
$E_0$ is the energy gap for particular QD, and the parameter $\alpha$ is  the linear dependence coefficient
between the squared line width and peak energy in the units of kT.  The values of  $\alpha$ are found to be
in the range from 0.48 to 0.63 for varied QDs at room temperature.

\begin{figure}
\centering
\begin{minipage}[h]{0.49\linewidth}
\center{\includegraphics[width=1\linewidth]{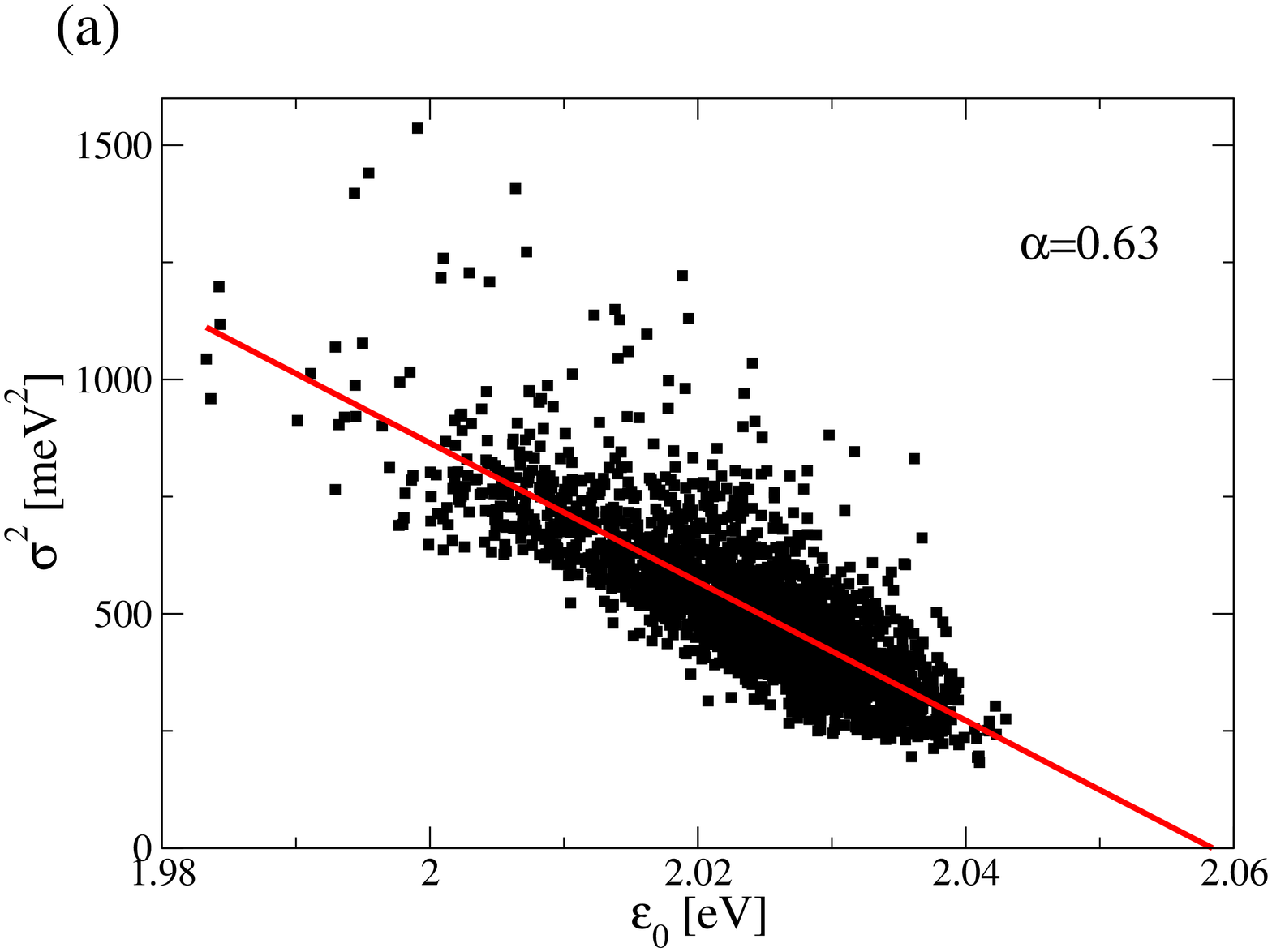}}
\end{minipage}
\hfill
\begin{minipage}[h]{0.49\linewidth}
\center{\includegraphics[width=1\linewidth]{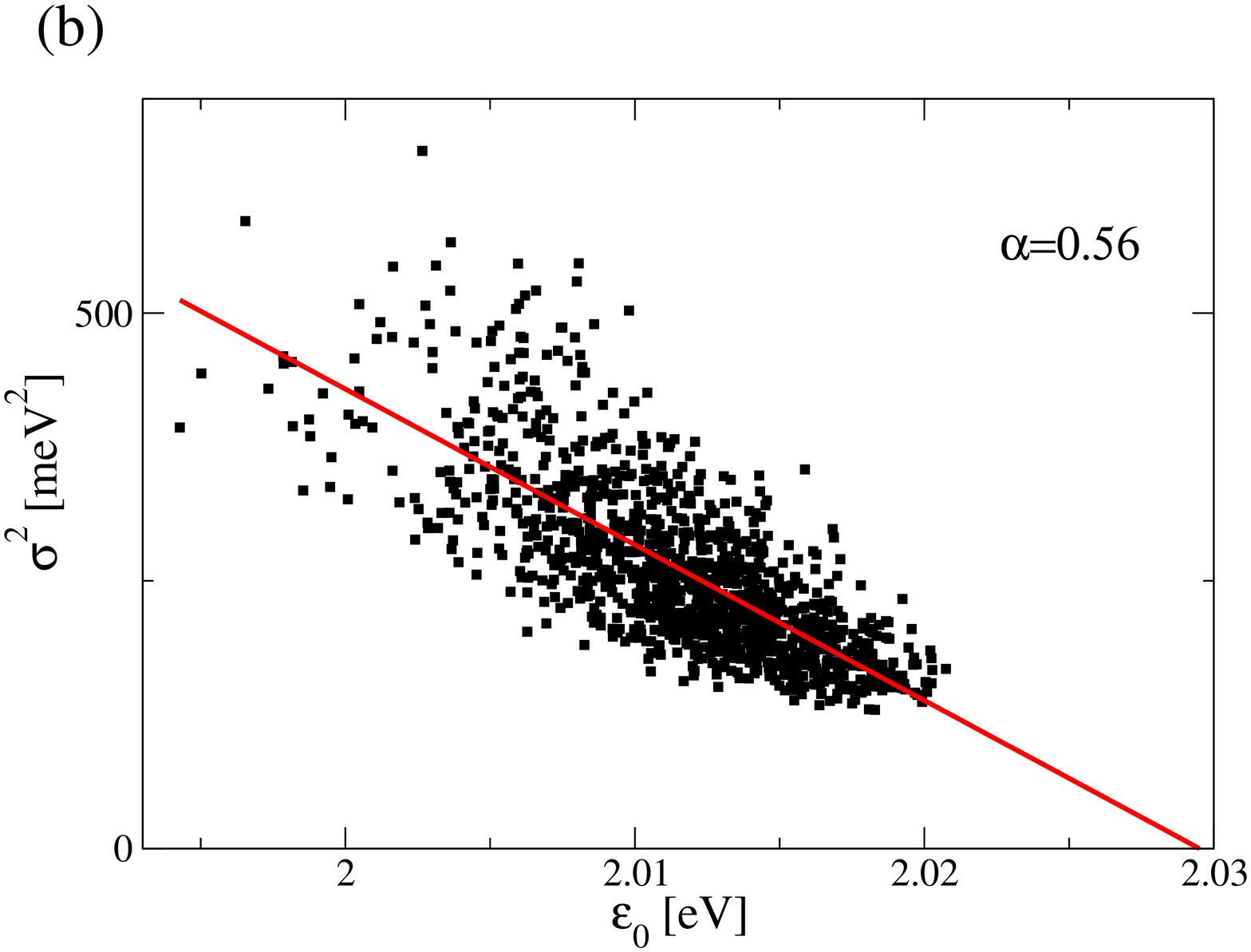}}
\end{minipage}
\vfill
\begin{minipage}[h]{0.49\linewidth}
\center{\includegraphics[width=1\linewidth]{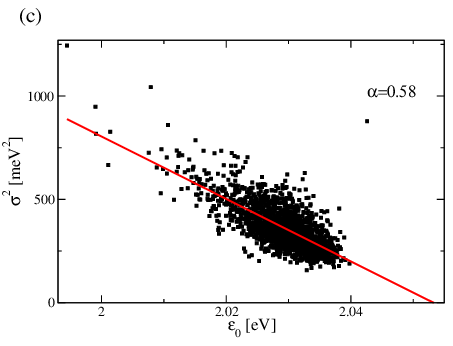}}
\end{minipage}
\hfill
\begin{minipage}[h]{0.49\linewidth}
\center{\includegraphics[width=1\linewidth]{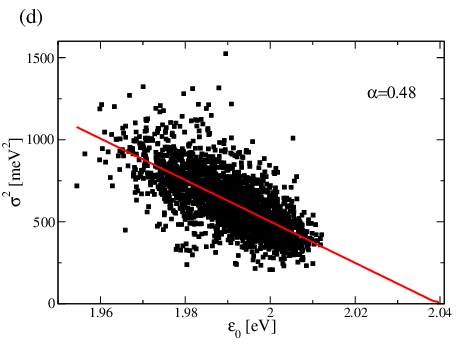}}
\end{minipage}
 \caption{The peak energy versus the linewidth squared for different single QDs (a) - (d) at room temperature (black points) and
 the linear fit Eq. (\ref{depend}) {with $T=300$ K}  (red lines).  The statistical error in the value of alpha in each fit is less then 0.014. }
\end{figure}

In order to characterize the shift of the spectra
 we found a squared peak energy displacement  of a typical single QD emission spectrum
$$D^2(\tau) =\left\langle (\epsilon_0(t)-\epsilon_0(t+\tau))^2\right \rangle$$
as a function of time $\tau$ following Ref. \cite{MulvaneyPRL2010}.
 As can be seen on Fig. 5, the averaged spectral shift squared $D^2$  is less than the $\sigma^2$ of a typical single QD spectrum
 for all delay times. But more importantly it is much smaller than linewidth squared when $\tau$ is equal to the signal accumulation time
  $\tau=200$ ms.  Thus, we can conclude that the observed linewidth is not related to the spectral shifts during this time period.

\begin{figure}
 \centering
 \includegraphics[width=\linewidth]{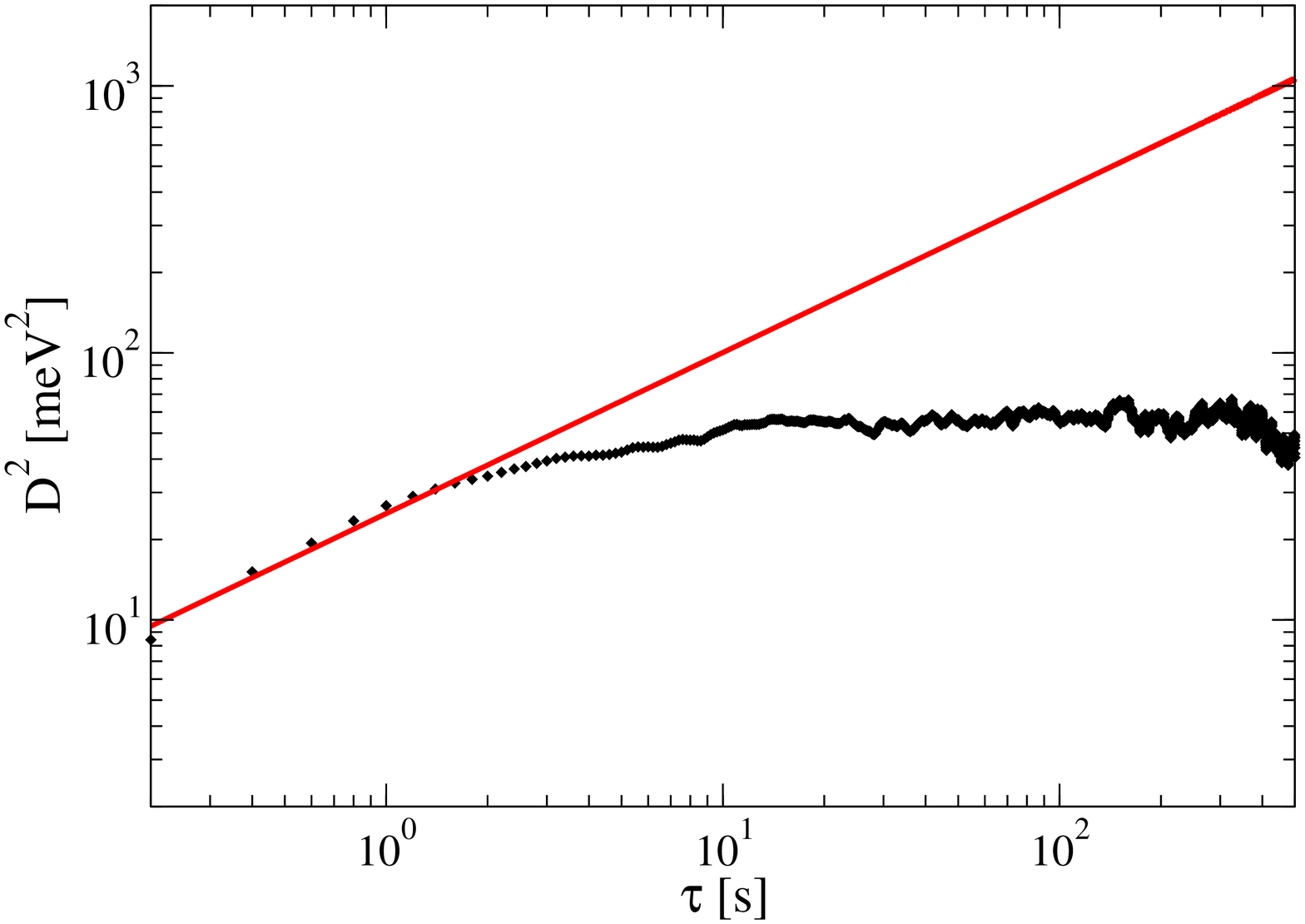}
 \caption{ Time dependence of $D^2 $ function for a single QD at room temperature (black diamonds). The red line is $\sim \tau^\beta$.
 The value of $\beta$ is 0.603}
\end{figure}

\section{Theory and Discussion}

Such a large value of the linewidth can be explained by multi-phonon excitation \cite{HuangPRSL1950,KuboPTP1955}.
 Let's consider the following
Hamiltonian of the QD electronic system interacting with $N$ phonon modes
\begin{equation}
 \hat{H} = \hat H_0 + A\sum\limits_{i=1}^N\hat{q}_i( a_i\ketbra{e}+b_i\ketbra{g})
 \label{H}
\end{equation}
 where
 \begin{equation}
 \hat H_0= \sum\limits_{i=1}^N \frac {\hat p_i^2}2 + \frac {\omega_i^2} 2   \hat{q}_i^2  + E_0 \ketbra{e}
\end{equation}
$E_0$ is the energy gap, $|g\rangle$ and  $|e\rangle$ are the ground state and excited electronic state of the  QD, respectively.
The parameter $A$ characterizes the electron-phonon interaction strength.
 $\hat q_i$ and $\hat p_i$ are the coordinate and momentum operators of  $i$-th phonon mode, characterized
by the frequency $\omega_i$. Both the excited state and the ground state are connected with the photon modes in the model.
The interaction of the excited state and the ground state with the $i$-th phonon is described by the dimensionless coefficients $a_i$ and $b_i$,
correspondingly.

The emission spectrum at a given value of $A$  has a Gaussian form with the following parameters (see details of the derivation in the Appendix):
\begin{equation}
\epsilon_{0}=E_0-A^2 \sum\limits_{i=1}^S \frac {a_i (a_i-b_i)}{\omega_i^2}
\label{epsilon0}
\end{equation}
\begin{equation}
\sigma^2 = kT A^2 \sum\limits_{i=1}^S \frac {(a_i-b_i)^2}{\omega_i^2}
\label{sigma2}
\end{equation}

Fluctuations of the linewidth within the model are explained by a slow variation of the parameter $A$.
 Such variations of the electron-phonon interaction  were observed experimentally  in single colloidal QDs
\cite{BawendiScience1997,WellerPRB2005} as well as in single chromoprotein molecules \cite{KohlerJPCB2012,KohlerAC2013}.

Variations of the parameter $A$ with time lead to shifts in the position of the maximum (spectral diffusion) correlated with the linewidth.
 Eqs. (\ref{epsilon0}-\ref{sigma2}) give the linear dependence Eq. (\ref{depend})
where
$$\alpha=\left[\sum\limits_{i=1}^N \frac {a_i (a_i-b_i)}{\omega_i^2}\right]^{-1} \sum\limits_{i=1}^N \frac {(a_i-b_i)^2}{\omega_i^2}$$
As seen on Fig. 4 the parameter $\alpha$ vary from one  QD to another, as well as $E_0$.
The model predictions are  consistent with the experimental results of Gomez et al. \cite{MulvaneyAPL2006}, since the electron-phonon interaction in QDs is not related to the dielectric properties of the environment.

In order to check the theory at various temperatures the spectroscopic experiment on one quantum dot was performed with heating and cooling of the sample. Twenty spectra were measured sequentially at each selected temperature in the range from 305.5 K to 353.6 K.
It was found that all the data can be well fitted by Eq.(\ref{depend})  provided the energy $E_0$ gap depends on temperature and
$\alpha$ keeps constant as seen  in Fig. 6. The $E_0$ value proves to decrease with temperature rise. The  changes in the effective band gap presumably occur due to thermal expansion. Importantly, this result doesn't depend on the process which led the system to that temperature (heating or cooling).
Note that this dependence of the "pure" energy gap $E_0$ on temperature is not due to the electron-phonon interaction
as is usually considered \cite{ChenAPL1991,TangPCCP2012,NaumovOS2019}.

\begin{figure}
 \centering
 \includegraphics[width=0.95\linewidth]{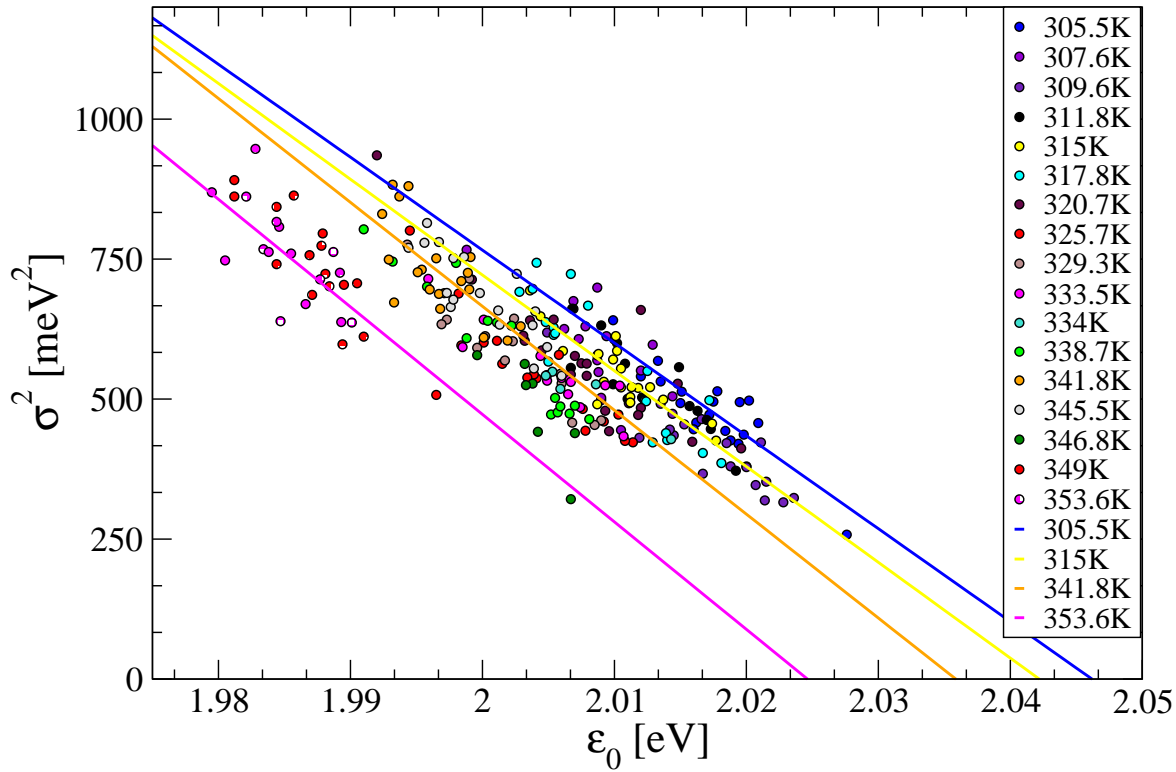}
 \caption{The peak energy versus the linewidth squared for a single QD at various temperatures (points).
  The solid lines indicate the theoretical prediction at four different temperatures. Parameter $\alpha=0.63$.}
\end{figure}

 Therefore, the  suggested model explains the fluctuations of the linewidth of a single QD emission at temperatures 300 K and above.
 But at the same time it predicts the spectral shifts correlated with the linewidth.

What is the mechanism of the variations in the magnitude of the electron-phonon interaction?
 The estimate shows that at a given excitation intensity of 100 W/cm$^2$, the average time between the absorption of photons of one QD is about one microsecond. Thus, the effect of multi-exciton states can be excluded from consideration. An increase in the temperature of a single QD after absorption of photons does not exceed 2 K, as estimated by Kuno et al. \cite{KunoJCP2001}, while thermal relaxation is of the order of 10 ps. This means that local heating can also be excluded from the possible causes of the phenomenon.
Plakhotnik et al. \cite{MulvaneyPRL2010} showed that the squared energy displacement of a single QD emission at cryogenic temperatures
has an anomalous (sublinear) behavior at short times  $D^2\sim \tau^\beta$, where $\beta<1$.
It was explained by introducing a number of stochastic two-level systems (TLS) having a wide distribution of flipping rates.
The squared energy displacement calculated with the use of our experimental data at room temperature also shows a similar sublinear time dependence (Fig.5).
That means that the TLS based  model can be used to describe the fluctuations of the electron-phonon interaction value
at high temperatures as well. A possible microscopic origin of the conformation change in the TLS could be due to the  jumps of the
surface or interface atom between two quasistable positions \cite{VoznyyPRL2014}.
Note the general interest regarding the microscopic nature of the SD processes which were observed for most single quantum emitters: single organic dye molecules
\cite{OrritScience1999,KadorPRL2003,SkinnerPRL1993}, single light harvesting complexes and proteins \cite{KohlerPRL2005}, color centers in diamonds \cite{BensonPRL2013}, single rare-earth ions in crystals \cite{SandoghdarNJPL2015}.
In many cases SD has been attributed to the tunneling processes in an emitter and/or its local surroundings. At the same time the relation between SD and phonon-assisted optical dephasing was always under discussion.

Empedocles and Bawendi \cite{BawendiScience1997} observed changes of the electron-phonon interaction parameter upon application of an external electric field.
It can be assumed that the shift of the peak of the spectrum in an external electric field is partially determined by a change in the electron - phonon interaction. To verify this assumption, additional experiments could be performed.

Single QD blinking also can be explained within a TLS based model of  \cite{FrantsuzovPRL2009}.
The Multiple Recombination Center (MRC) model  suggested by Frantsuzov et al. \cite{FrantsuzovPRL2009} reproduces the key properties of single QD blinking, such as the ON and OFF time distribution functions \cite{FrantsuzovPRL2009}, the power spectral density \cite{FrantsuzovNL2013},  and the long-term correlations between subsequent blinking times \cite{VolkanNL2010}.
The similarity in temporal fluctuations of the spectrum and the emission intensity of a single QD allows one to make the assumption that both phenomena can be explained by a unified mechanism \cite{BusovOS2019}.

In conclusion, our experiments show a linear correlation between the position of the maximum and the linewidth squared of a single QD emission spectrum at room temperature and above.
 In order to explain the experimental results, we consider a model of QD emission spectrum linewidth fluctuations based on a slow variation of the electron-phonon interaction. The model was tested using the data of a unique single QD spectroscopy experiment under heating and cooling conditions.

\section* {Acknowledgements}

The study was supported by the Russian Foundation for Basic Research, project 16-02-00713.
The measurements were  carried out under  the State Contract of the Institute of Spectroscopy RAS.
 Luminescence Microscopy Technique with detection of Single Quantum Dots with Nanometer Spatial Resolution is developed under support of Russian Science Foundation (project 17-72-20266, head I.Yu Eremchev).

\section* {Appendix: DERIVATION OF EQUATIONS (4) AND (5)}

Potential energy of the the excited electronic state is given by the following formula:
$$U_e(q)=\sum\limits_{i=1}^N \left(\frac {\omega_i^2} 2 q_i^2 + A{q}_i a_i\right)$$
In the classical limit $\hbar \omega_i \ll kT$ the probability distribution function of the coordinates is given by the Boltzmann distribution:
\begin{equation}
 P(q)=\frac 1 Z \exp\left\{-\frac 1 {kT} \sum\limits_{i=1}^N \left(\frac {\omega_i^2} 2 q_i^2 + A {q}_i a_i\right)\right\}
\label{Distr}
\end{equation}
where $Z$ is the partition function
$$Z=\int dq_1 \int dq_2\cdots\int dq_N\,\exp\left\{-\frac 1 {kT} \sum\limits_{i=1}^N \left(\frac {\omega_i^2} 2 q_i^2 + A {q}_i a_i\right)\right\}$$
It is assumed that the thermal relaxation is much faster than the variations of the parameter $A$.
The energy of the emitted photon at given values of the phonon coordinates is equal to the difference between the energies of the excited and ground states
\begin{equation}
 \epsilon= E_0+A\sum\limits_{i=1}^N  (a_i-b_i)q_i
 \label{epsilon}
\end{equation}

Eq.(\ref{Distr}) is a multi-dimension Gaussian distribution, Eq.(\ref{epsilon}) is a linear function of the coordinates $q_i$.
It follows that the distribution of $\epsilon$ is also Gaussian
$$ p(\epsilon)=\frac 1 {\sqrt{2\pi}\sigma}\exp\left\{-\frac{(\epsilon-\epsilon_0)^2}{2\sigma^2}\right\}$$
where the parameters $\epsilon_0$ and $\sigma^2$ can be found by averaging over the distribution (\ref{Distr})
\begin{equation}
\epsilon_0= \bar \epsilon  = E_0+ A\sum\limits_{i=1}^N (a_i-b_i) \langle q_i\rangle
 \label{Avepsilon}
\end{equation}
\begin{equation}
\sigma^2=\langle (\epsilon-\bar \epsilon )^2 \rangle = A^2 \sum\limits_{i=1}^N (a_i-b_i)^2 \left\langle \left(q_i-\bar q_i \right)^2\right \rangle
 \label{Avsigma}
\end{equation}
The mean values for the coordinates can be easily found by integration over distribution (\ref{Distr})
$$\bar q_i=-A\frac {a_i}{\omega_i^2}$$

$$\left\langle \left(q_i- \bar q_i\right)^2 \right\rangle=\frac {kT}{\omega_i^2}$$

Substituting these expressions into the Eqs. (\ref{Avepsilon}-\ref{Avsigma}) gives  Eqs. (\ref{epsilon0}-\ref{sigma2}).

\bibliography{Spectral}

\begin{thebibliography}{34}%
\makeatletter
\providecommand \@ifxundefined [1]{%
 \@ifx{#1\undefined}
}%
\providecommand \@ifnum [1]{%
 \ifnum #1\expandafter \@firstoftwo
 \else \expandafter \@secondoftwo
 \fi
}%
\providecommand \@ifx [1]{%
 \ifx #1\expandafter \@firstoftwo
 \else \expandafter \@secondoftwo
 \fi
}%
\providecommand \natexlab [1]{#1}%
\providecommand \enquote  [1]{``#1''}%
\providecommand \bibnamefont  [1]{#1}%
\providecommand \bibfnamefont [1]{#1}%
\providecommand \citenamefont [1]{#1}%
\providecommand \href@noop [0]{\@secondoftwo}%
\providecommand \href [0]{\begingroup \@sanitize@url \@href}%
\providecommand \@href[1]{\@@startlink{#1}\@@href}%
\providecommand \@@href[1]{\endgroup#1\@@endlink}%
\providecommand \@sanitize@url [0]{\catcode `\\12\catcode `\$12\catcode
  `\&12\catcode `\#12\catcode `\^12\catcode `\_12\catcode `\%12\relax}%
\providecommand \@@startlink[1]{}%
\providecommand \@@endlink[0]{}%
\providecommand \url  [0]{\begingroup\@sanitize@url \@url }%
\providecommand \@url [1]{\endgroup\@href {#1}{\urlprefix }}%
\providecommand \urlprefix  [0]{URL }%
\providecommand \Eprint [0]{\href }%
\providecommand \doibase [0]{http://dx.doi.org/}%
\providecommand \selectlanguage [0]{\@gobble}%
\providecommand \bibinfo  [0]{\@secondoftwo}%
\providecommand \bibfield  [0]{\@secondoftwo}%
\providecommand \translation [1]{[#1]}%
\providecommand \BibitemOpen [0]{}%
\providecommand \bibitemStop [0]{}%
\providecommand \bibitemNoStop [0]{.\EOS\space}%
\providecommand \EOS [0]{\spacefactor3000\relax}%
\providecommand \BibitemShut  [1]{\csname bibitem#1\endcsname}%
\let\auto@bib@innerbib\@empty
\bibitem [{\citenamefont {Nirmal}\ \emph {et~al.}(1996)\citenamefont {Nirmal},
  \citenamefont {Dabbousi}, \citenamefont {G.}, \citenamefont {Macklin},
  \citenamefont {Trautman}, \citenamefont {Harris},\ and\ \citenamefont
  {Brus}}]{BrusNature1996}%
  \BibitemOpen
  \bibfield  {author} {\bibinfo {author} {\bibfnamefont {M.}~\bibnamefont
  {Nirmal}}, \bibinfo {author} {\bibfnamefont {B.~O.}\ \bibnamefont
  {Dabbousi}}, \bibinfo {author} {\bibfnamefont {B.~M.}\ \bibnamefont {G.}},
  \bibinfo {author} {\bibfnamefont {J.~J.}\ \bibnamefont {Macklin}}, \bibinfo
  {author} {\bibfnamefont {J.~K.}\ \bibnamefont {Trautman}}, \bibinfo {author}
  {\bibfnamefont {T.~D.}\ \bibnamefont {Harris}}, \ and\ \bibinfo {author}
  {\bibfnamefont {L.~E.}\ \bibnamefont {Brus}},\ }\href@noop {} {\bibfield
  {journal} {\bibinfo  {journal} {Nature}\ }\textbf {\bibinfo {volume} {383}},\
  \bibinfo {pages} {802} (\bibinfo {year} {1996})}\BibitemShut {NoStop}%
\bibitem [{\citenamefont {Empedocles}\ \emph {et~al.}(1996)\citenamefont
  {Empedocles}, \citenamefont {Norris},\ and\ \citenamefont
  {Bawendi}}]{BawendiPRL1996}%
  \BibitemOpen
  \bibfield  {author} {\bibinfo {author} {\bibfnamefont {S.~A.}\ \bibnamefont
  {Empedocles}}, \bibinfo {author} {\bibfnamefont {D.~J.}\ \bibnamefont
  {Norris}}, \ and\ \bibinfo {author} {\bibfnamefont {M.~G.}\ \bibnamefont
  {Bawendi}},\ }\href@noop {} {\bibfield  {journal} {\bibinfo  {journal} {Phys.
  Rev. Lett.}\ }\textbf {\bibinfo {volume} {77}},\ \bibinfo {pages} {3873}
  (\bibinfo {year} {1996})}\BibitemShut {NoStop}%
\bibitem [{\citenamefont {Blanton}\ \emph {et~al.}(1996)\citenamefont
  {Blanton}, \citenamefont {Hines},\ and\ \citenamefont
  {Guyot-Sionnest}}]{SionnestAPL1996}%
  \BibitemOpen
  \bibfield  {author} {\bibinfo {author} {\bibfnamefont {S.~A.}\ \bibnamefont
  {Blanton}}, \bibinfo {author} {\bibfnamefont {M.~A.}\ \bibnamefont {Hines}},
  \ and\ \bibinfo {author} {\bibfnamefont {P.}~\bibnamefont {Guyot-Sionnest}},\
  }\href@noop {} {\bibfield  {journal} {\bibinfo  {journal} {App. Phys. Lett.}\
  }\textbf {\bibinfo {volume} {69}},\ \bibinfo {pages} {3905 } (\bibinfo {year}
  {1996})}\BibitemShut {NoStop}%
\bibitem [{\citenamefont {Beyler}\ \emph {et~al.}(2013)\citenamefont {Beyler},
  \citenamefont {Marshall}, \citenamefont {Cui}, \citenamefont {Brokmann},\
  and\ \citenamefont {Bawendi}}]{BawendiPRL2013}%
  \BibitemOpen
  \bibfield  {author} {\bibinfo {author} {\bibfnamefont {A.~P.}\ \bibnamefont
  {Beyler}}, \bibinfo {author} {\bibfnamefont {L.~F.}\ \bibnamefont
  {Marshall}}, \bibinfo {author} {\bibfnamefont {J.}~\bibnamefont {Cui}},
  \bibinfo {author} {\bibfnamefont {X.}~\bibnamefont {Brokmann}}, \ and\
  \bibinfo {author} {\bibfnamefont {M.~G.}\ \bibnamefont {Bawendi}},\
  }\href@noop {} {\bibfield  {journal} {\bibinfo  {journal} {Phys. Rev. Lett.}\
  }\textbf {\bibinfo {volume} {111}},\ \bibinfo {pages} {177401} (\bibinfo
  {year} {2013})}\BibitemShut {NoStop}%
\bibitem [{\citenamefont {Empedocles}\ and\ \citenamefont
  {Bawendi}(1999)}]{BawendiJPCB1999}%
  \BibitemOpen
  \bibfield  {author} {\bibinfo {author} {\bibfnamefont {S.~A.}\ \bibnamefont
  {Empedocles}}\ and\ \bibinfo {author} {\bibfnamefont {M.~G.}\ \bibnamefont
  {Bawendi}},\ }\href@noop {} {\bibfield  {journal} {\bibinfo  {journal} {J.
  Phys. Chem. B}\ }\textbf {\bibinfo {volume} {103}},\ \bibinfo {pages} {1826}
  (\bibinfo {year} {1999})}\BibitemShut {NoStop}%
\bibitem [{\citenamefont {Empedocles}\ and\ \citenamefont
  {Bawendi}(1997)}]{BawendiScience1997}%
  \BibitemOpen
  \bibfield  {author} {\bibinfo {author} {\bibfnamefont {S.~A.}\ \bibnamefont
  {Empedocles}}\ and\ \bibinfo {author} {\bibfnamefont {M.~G.}\ \bibnamefont
  {Bawendi}},\ }\href@noop {} {\bibfield  {journal} {\bibinfo  {journal}
  {Science}\ }\textbf {\bibinfo {volume} {278}},\ \bibinfo {pages} {2114}
  (\bibinfo {year} {1997})}\BibitemShut {NoStop}%
\bibitem [{\citenamefont {Muller}\ \emph {et~al.}(2004)\citenamefont {Muller},
  \citenamefont {Lupton}, \citenamefont {Rogach}, \citenamefont {Feldmann},
  \citenamefont {Talapin},\ and\ \citenamefont {H.Weller}}]{WellerPRL2004}%
  \BibitemOpen
  \bibfield  {author} {\bibinfo {author} {\bibfnamefont {J.}~\bibnamefont
  {Muller}}, \bibinfo {author} {\bibfnamefont {J.~M.}\ \bibnamefont {Lupton}},
  \bibinfo {author} {\bibfnamefont {A.~L.}\ \bibnamefont {Rogach}}, \bibinfo
  {author} {\bibfnamefont {J.}~\bibnamefont {Feldmann}}, \bibinfo {author}
  {\bibfnamefont {D.~V.}\ \bibnamefont {Talapin}}, \ and\ \bibinfo {author}
  {\bibnamefont {H.Weller}},\ }\href@noop {} {\bibfield  {journal} {\bibinfo
  {journal} {Phys. Rev. Lett.}\ }\textbf {\bibinfo {volume} {93}},\ \bibinfo
  {pages} {167402} (\bibinfo {year} {2004})}\BibitemShut {NoStop}%
\bibitem [{\citenamefont {Muller}\ \emph {et~al.}(2005)\citenamefont {Muller},
  \citenamefont {Lupton}, \citenamefont {Rogach}, \citenamefont {Feldmann},
  \citenamefont {Talapin},\ and\ \citenamefont {H.Weller}}]{WellerPRB2005}%
  \BibitemOpen
  \bibfield  {author} {\bibinfo {author} {\bibfnamefont {J.}~\bibnamefont
  {Muller}}, \bibinfo {author} {\bibfnamefont {J.~M.}\ \bibnamefont {Lupton}},
  \bibinfo {author} {\bibfnamefont {A.~L.}\ \bibnamefont {Rogach}}, \bibinfo
  {author} {\bibfnamefont {J.}~\bibnamefont {Feldmann}}, \bibinfo {author}
  {\bibfnamefont {D.~V.}\ \bibnamefont {Talapin}}, \ and\ \bibinfo {author}
  {\bibnamefont {H.Weller}},\ }\href@noop {} {\bibfield  {journal} {\bibinfo
  {journal} {Phys. Rev. B}\ }\textbf {\bibinfo {volume} {72}},\ \bibinfo
  {pages} {205339} (\bibinfo {year} {2005})}\BibitemShut {NoStop}%
\bibitem [{\citenamefont {Gomez}\ \emph {et~al.}(2006)\citenamefont {Gomez},
  \citenamefont {van Embden},\ and\ \citenamefont
  {Mulvaney}}]{MulvaneyAPL2006}%
  \BibitemOpen
  \bibfield  {author} {\bibinfo {author} {\bibfnamefont {D.~E.}\ \bibnamefont
  {Gomez}}, \bibinfo {author} {\bibfnamefont {J.}~\bibnamefont {van Embden}}, \
  and\ \bibinfo {author} {\bibfnamefont {P.}~\bibnamefont {Mulvaney}},\
  }\href@noop {} {\bibfield  {journal} {\bibinfo  {journal} {Appl. Phys.
  Lett.}\ }\textbf {\bibinfo {volume} {88}},\ \bibinfo {pages} {154106}
  (\bibinfo {year} {2006})}\BibitemShut {NoStop}%
\bibitem [{\citenamefont {Dzhagan}\ \emph {et~al.}(2018)\citenamefont
  {Dzhagan}, \citenamefont {Azhniuk}, \citenamefont {A},\ and\ \citenamefont
  {Zahn}}]{DzhaganJPD2018}%
  \BibitemOpen
  \bibfield  {author} {\bibinfo {author} {\bibfnamefont {V.~M.}\ \bibnamefont
  {Dzhagan}}, \bibinfo {author} {\bibfnamefont {Y.~M.}\ \bibnamefont
  {Azhniuk}}, \bibinfo {author} {\bibfnamefont {G.~M.}\ \bibnamefont {A}}, \
  and\ \bibinfo {author} {\bibfnamefont {D.~R.~T.}\ \bibnamefont {Zahn}},\
  }\href@noop {} {\bibfield  {journal} {\bibinfo  {journal} {J. Phys. D}\
  }\textbf {\bibinfo {volume} {51}},\ \bibinfo {pages} {503001} (\bibinfo
  {year} {2018})}\BibitemShut {NoStop}%
\bibitem [{\citenamefont {Huang}\ and\ \citenamefont
  {Rhys}(1950)}]{HuangPRSL1950}%
  \BibitemOpen
  \bibfield  {author} {\bibinfo {author} {\bibfnamefont {K.}~\bibnamefont
  {Huang}}\ and\ \bibinfo {author} {\bibfnamefont {A.}~\bibnamefont {Rhys}},\
  }\href@noop {} {\bibfield  {journal} {\bibinfo  {journal} {Proc. Royal Soc.
  Lond. A, Math. Phys.}\ }\textbf {\bibinfo {volume} {204}},\ \bibinfo {pages}
  {406 } (\bibinfo {year} {1950})}\BibitemShut {NoStop}%
\bibitem [{\citenamefont {Kubo}\ and\ \citenamefont
  {Toyozawa}(1955)}]{KuboPTP1955}%
  \BibitemOpen
  \bibfield  {author} {\bibinfo {author} {\bibfnamefont {R.}~\bibnamefont
  {Kubo}}\ and\ \bibinfo {author} {\bibfnamefont {Y.}~\bibnamefont
  {Toyozawa}},\ }\href@noop {} {\bibfield  {journal} {\bibinfo  {journal}
  {Prog. Theor. Phys.}\ }\textbf {\bibinfo {volume} {13}},\ \bibinfo {pages}
  {160 } (\bibinfo {year} {1955})}\BibitemShut {NoStop}%
\bibitem [{\citenamefont {Eremchev}\ \emph {et~al.}(2019)\citenamefont
  {Eremchev}, \citenamefont {Eremchev},\ and\ \citenamefont
  {Naumov}}]{NaumovUP2018}%
  \BibitemOpen
  \bibfield  {author} {\bibinfo {author} {\bibfnamefont {I.~Y.}\ \bibnamefont
  {Eremchev}}, \bibinfo {author} {\bibfnamefont {M.~Y.}\ \bibnamefont
  {Eremchev}}, \ and\ \bibinfo {author} {\bibfnamefont {A.~V.}\ \bibnamefont
  {Naumov}},\ }\href@noop {} {\bibfield  {journal} {\bibinfo  {journal} {Phys.
  Usp.}\ }\textbf {\bibinfo {volume} {62}},\ \bibinfo {pages} {294 } (\bibinfo
  {year} {2019})}\BibitemShut {NoStop}%
\bibitem [{\citenamefont {Eremchev}\ \emph {et~al.}(2009)\citenamefont
  {Eremchev}, \citenamefont {Naumov}, \citenamefont {Vainer},\ and\
  \citenamefont {Kador}}]{EremchevJCP2009}%
  \BibitemOpen
  \bibfield  {author} {\bibinfo {author} {\bibfnamefont {I.~Y.}\ \bibnamefont
  {Eremchev}}, \bibinfo {author} {\bibfnamefont {A.~V.}\ \bibnamefont
  {Naumov}}, \bibinfo {author} {\bibfnamefont {Y.~G.}\ \bibnamefont {Vainer}},
  \ and\ \bibinfo {author} {\bibfnamefont {L.}~\bibnamefont {Kador}},\
  }\href@noop {} {\bibfield  {journal} {\bibinfo  {journal} {J. Chem. Phys.}\
  }\textbf {\bibinfo {volume} {130}},\ \bibinfo {pages} {184507} (\bibinfo
  {year} {2009})}\BibitemShut {NoStop}%
\bibitem [{\citenamefont {Karimullin}\ and\ \citenamefont
  {Naumov}(2014)}]{KarimulinJL2014}%
  \BibitemOpen
  \bibfield  {author} {\bibinfo {author} {\bibfnamefont {K.~R.}\ \bibnamefont
  {Karimullin}}\ and\ \bibinfo {author} {\bibfnamefont {A.~V.}\ \bibnamefont
  {Naumov}},\ }\href@noop {} {\bibfield  {journal} {\bibinfo  {journal} {J.
  Lumin.}\ }\textbf {\bibinfo {volume} {152}},\ \bibinfo {pages} {15 }
  (\bibinfo {year} {2014})}\BibitemShut {NoStop}%
\bibitem [{\citenamefont {Eremchev}\ \emph {et~al.}(2018)\citenamefont
  {Eremchev}, \citenamefont {Lozing}, \citenamefont {Baev}, \citenamefont
  {Tarasevich}, \citenamefont {Gladush}, \citenamefont {Rozhentsov},\ and\
  \citenamefont {Naumov}}]{NaumovJL2018}%
  \BibitemOpen
  \bibfield  {author} {\bibinfo {author} {\bibfnamefont {I.~Y.}\ \bibnamefont
  {Eremchev}}, \bibinfo {author} {\bibfnamefont {N.~A.}\ \bibnamefont
  {Lozing}}, \bibinfo {author} {\bibfnamefont {A.~A.}\ \bibnamefont {Baev}},
  \bibinfo {author} {\bibfnamefont {A.~O.}\ \bibnamefont {Tarasevich}},
  \bibinfo {author} {\bibfnamefont {M.~G.}\ \bibnamefont {Gladush}}, \bibinfo
  {author} {\bibfnamefont {A.~A.}\ \bibnamefont {Rozhentsov}}, \ and\ \bibinfo
  {author} {\bibfnamefont {A.}~\bibnamefont {Naumov}},\ }\href@noop {}
  {\bibfield  {journal} {\bibinfo  {journal} {JETP Letters}\ }\textbf {\bibinfo
  {volume} {108}},\ \bibinfo {pages} {30 } (\bibinfo {year}
  {2018})}\BibitemShut {NoStop}%
\bibitem [{\citenamefont {Plakhotnik}\ \emph {et~al.}(2010)\citenamefont
  {Plakhotnik}, \citenamefont {Fernee}, \citenamefont {Littleton},
  \citenamefont {Rubinsztein-Dunlop}, \citenamefont {Potzner},\ and\
  \citenamefont {Mulvaney}}]{MulvaneyPRL2010}%
  \BibitemOpen
  \bibfield  {author} {\bibinfo {author} {\bibfnamefont {T.}~\bibnamefont
  {Plakhotnik}}, \bibinfo {author} {\bibfnamefont {M.~J.}\ \bibnamefont
  {Fernee}}, \bibinfo {author} {\bibfnamefont {B.}~\bibnamefont {Littleton}},
  \bibinfo {author} {\bibfnamefont {H.}~\bibnamefont {Rubinsztein-Dunlop}},
  \bibinfo {author} {\bibfnamefont {C.}~\bibnamefont {Potzner}}, \ and\
  \bibinfo {author} {\bibfnamefont {P.}~\bibnamefont {Mulvaney}},\ }\href@noop
  {} {\bibfield  {journal} {\bibinfo  {journal} {Phys. Rev. Lett.}\ }\textbf
  {\bibinfo {volume} {105}},\ \bibinfo {pages} {167402} (\bibinfo {year}
  {2010})}\BibitemShut {NoStop}%
\bibitem [{\citenamefont {Kunz}\ \emph {et~al.}(2012)\citenamefont {Kunz},
  \citenamefont {Timpmann}, \citenamefont {Southall}, \citenamefont {Cogdell},
  \citenamefont {Freiberg},\ and\ \citenamefont {Kohler}}]{KohlerJPCB2012}%
  \BibitemOpen
  \bibfield  {author} {\bibinfo {author} {\bibfnamefont {R.}~\bibnamefont
  {Kunz}}, \bibinfo {author} {\bibfnamefont {K.}~\bibnamefont {Timpmann}},
  \bibinfo {author} {\bibfnamefont {J.}~\bibnamefont {Southall}}, \bibinfo
  {author} {\bibfnamefont {R.~J.}\ \bibnamefont {Cogdell}}, \bibinfo {author}
  {\bibfnamefont {A.}~\bibnamefont {Freiberg}}, \ and\ \bibinfo {author}
  {\bibfnamefont {J.}~\bibnamefont {Kohler}},\ }\href@noop {} {\bibfield
  {journal} {\bibinfo  {journal} {J. Phys. Chem. B}\ }\textbf {\bibinfo
  {volume} {116}},\ \bibinfo {pages} {11017 } (\bibinfo {year}
  {2012})}\BibitemShut {NoStop}%
\bibitem [{\citenamefont {Kunz}\ \emph {et~al.}(2013)\citenamefont {Kunz},
  \citenamefont {Timpmann}, \citenamefont {Southall}, \citenamefont {Cogdell},
  \citenamefont {Freiberg},\ and\ \citenamefont {Kohler}}]{KohlerAC2013}%
  \BibitemOpen
  \bibfield  {author} {\bibinfo {author} {\bibfnamefont {R.}~\bibnamefont
  {Kunz}}, \bibinfo {author} {\bibfnamefont {K.}~\bibnamefont {Timpmann}},
  \bibinfo {author} {\bibfnamefont {J.}~\bibnamefont {Southall}}, \bibinfo
  {author} {\bibfnamefont {R.~J.}\ \bibnamefont {Cogdell}}, \bibinfo {author}
  {\bibfnamefont {A.}~\bibnamefont {Freiberg}}, \ and\ \bibinfo {author}
  {\bibfnamefont {J.}~\bibnamefont {Kohler}},\ }\href@noop {} {\bibfield
  {journal} {\bibinfo  {journal} {Angew. Chem. Int. Ed.}\ }\textbf {\bibinfo
  {volume} {52}},\ \bibinfo {pages} {8726 } (\bibinfo {year}
  {2013})}\BibitemShut {NoStop}%
\bibitem [{\citenamefont {O'Donnell}\ and\ \citenamefont
  {Chen}(1991)}]{ChenAPL1991}%
  \BibitemOpen
  \bibfield  {author} {\bibinfo {author} {\bibfnamefont {K.~P.}\ \bibnamefont
  {O'Donnell}}\ and\ \bibinfo {author} {\bibfnamefont {X.}~\bibnamefont
  {Chen}},\ }\href@noop {} {\bibfield  {journal} {\bibinfo  {journal} {Appl.
  Phys. Lett.}\ }\textbf {\bibinfo {volume} {58}},\ \bibinfo {pages} {2924}
  (\bibinfo {year} {1991})}\BibitemShut {NoStop}%
\bibitem [{\citenamefont {Wen}\ \emph {et~al.}(2012)\citenamefont {Wen},
  \citenamefont {Sitt}, \citenamefont {Yu}, \citenamefont {Toha},\ and\
  \citenamefont {Tang}}]{TangPCCP2012}%
  \BibitemOpen
  \bibfield  {author} {\bibinfo {author} {\bibfnamefont {X.}~\bibnamefont
  {Wen}}, \bibinfo {author} {\bibfnamefont {A.}~\bibnamefont {Sitt}}, \bibinfo
  {author} {\bibfnamefont {P.}~\bibnamefont {Yu}}, \bibinfo {author}
  {\bibfnamefont {Y.-R.}\ \bibnamefont {Toha}}, \ and\ \bibinfo {author}
  {\bibfnamefont {J.}~\bibnamefont {Tang}},\ }\href@noop {} {\bibfield
  {journal} {\bibinfo  {journal} {Phys. Chem. Chem. Phys.}\ }\textbf {\bibinfo
  {volume} {14}},\ \bibinfo {pages} {3505 } (\bibinfo {year}
  {2012})}\BibitemShut {NoStop}%
\bibitem [{\citenamefont {Magaryan}\ \emph {et~al.}(2019)\citenamefont
  {Magaryan}, \citenamefont {Karimullin}, \citenamefont {Vasil'eva},\ and\
  \citenamefont {Naumov}}]{NaumovOS2019}%
  \BibitemOpen
  \bibfield  {author} {\bibinfo {author} {\bibfnamefont {K.~A.}\ \bibnamefont
  {Magaryan}}, \bibinfo {author} {\bibfnamefont {K.~R.}\ \bibnamefont
  {Karimullin}}, \bibinfo {author} {\bibfnamefont {I.~A.}\ \bibnamefont
  {Vasil'eva}}, \ and\ \bibinfo {author} {\bibfnamefont {A.~V.}\ \bibnamefont
  {Naumov}},\ }\href@noop {} {\bibfield  {journal} {\bibinfo  {journal} {Opt.
  Spectr.}\ }\textbf {\bibinfo {volume} {126}},\ \bibinfo {pages} {41 }
  (\bibinfo {year} {2019})}\BibitemShut {NoStop}%
\bibitem [{\citenamefont {Kuno}\ \emph {et~al.}(2001)\citenamefont {Kuno},
  \citenamefont {Fromm}, \citenamefont {Hammann}, \citenamefont {Gallagher},\
  and\ \citenamefont {Nesbitt}}]{KunoJCP2001}%
  \BibitemOpen
  \bibfield  {author} {\bibinfo {author} {\bibfnamefont {M.}~\bibnamefont
  {Kuno}}, \bibinfo {author} {\bibfnamefont {D.~P.}\ \bibnamefont {Fromm}},
  \bibinfo {author} {\bibfnamefont {H.~F.}\ \bibnamefont {Hammann}}, \bibinfo
  {author} {\bibfnamefont {A.}~\bibnamefont {Gallagher}}, \ and\ \bibinfo
  {author} {\bibfnamefont {D.~J.}\ \bibnamefont {Nesbitt}},\ }\href@noop {}
  {\bibfield  {journal} {\bibinfo  {journal} {J. Chem. Phys.}\ }\textbf
  {\bibinfo {volume} {115}},\ \bibinfo {pages} {1028} (\bibinfo {year}
  {2001})}\BibitemShut {NoStop}%
\bibitem [{\citenamefont {Voznyy}\ and\ \citenamefont
  {Sargent}(2014)}]{VoznyyPRL2014}%
  \BibitemOpen
  \bibfield  {author} {\bibinfo {author} {\bibfnamefont {O.}~\bibnamefont
  {Voznyy}}\ and\ \bibinfo {author} {\bibfnamefont {E.~H.}\ \bibnamefont
  {Sargent}},\ }\href@noop {} {\bibfield  {journal} {\bibinfo  {journal} {Phys.
  Rev. Lett.}\ }\textbf {\bibinfo {volume} {112}},\ \bibinfo {pages} {157401}
  (\bibinfo {year} {2014})}\BibitemShut {NoStop}%
\bibitem [{\citenamefont {Moerner}\ and\ \citenamefont
  {Orrit}(1999)}]{OrritScience1999}%
  \BibitemOpen
  \bibfield  {author} {\bibinfo {author} {\bibfnamefont {W.~E.}\ \bibnamefont
  {Moerner}}\ and\ \bibinfo {author} {\bibfnamefont {M.}~\bibnamefont
  {Orrit}},\ }\href@noop {} {\bibfield  {journal} {\bibinfo  {journal}
  {Science}\ }\textbf {\bibinfo {volume} {283}},\ \bibinfo {pages} {1670}
  (\bibinfo {year} {1999})}\BibitemShut {NoStop}%
\bibitem [{\citenamefont {Barkai}\ \emph {et~al.}(2003)\citenamefont {Barkai},
  \citenamefont {Naumov}, \citenamefont {Vainer}, \citenamefont {Bauer},\ and\
  \citenamefont {Kador}}]{KadorPRL2003}%
  \BibitemOpen
  \bibfield  {author} {\bibinfo {author} {\bibfnamefont {E.}~\bibnamefont
  {Barkai}}, \bibinfo {author} {\bibfnamefont {A.~V.}\ \bibnamefont {Naumov}},
  \bibinfo {author} {\bibfnamefont {Y.~G.}\ \bibnamefont {Vainer}}, \bibinfo
  {author} {\bibfnamefont {M.}~\bibnamefont {Bauer}}, \ and\ \bibinfo {author}
  {\bibfnamefont {L.}~\bibnamefont {Kador}},\ }\href@noop {} {\bibfield
  {journal} {\bibinfo  {journal} {Phys. Rev. Lett.}\ }\textbf {\bibinfo
  {volume} {91}},\ \bibinfo {pages} {075502} (\bibinfo {year}
  {2003})}\BibitemShut {NoStop}%
\bibitem [{\citenamefont {Reilly}\ and\ \citenamefont
  {Skinner}(1993)}]{SkinnerPRL1993}%
  \BibitemOpen
  \bibfield  {author} {\bibinfo {author} {\bibfnamefont {P.~D.}\ \bibnamefont
  {Reilly}}\ and\ \bibinfo {author} {\bibfnamefont {J.~L.}\ \bibnamefont
  {Skinner}},\ }\href@noop {} {\bibfield  {journal} {\bibinfo  {journal} {Phys.
  Rev. Lett.}\ }\textbf {\bibinfo {volume} {71}},\ \bibinfo {pages} {4257}
  (\bibinfo {year} {1993})}\BibitemShut {NoStop}%
\bibitem [{\citenamefont {Hofmann}\ \emph {et~al.}(2005)\citenamefont
  {Hofmann}, \citenamefont {Michel}, \citenamefont {van Heel},\ and\
  \citenamefont {Kohler}}]{KohlerPRL2005}%
  \BibitemOpen
  \bibfield  {author} {\bibinfo {author} {\bibfnamefont {C.}~\bibnamefont
  {Hofmann}}, \bibinfo {author} {\bibfnamefont {H.}~\bibnamefont {Michel}},
  \bibinfo {author} {\bibfnamefont {M.}~\bibnamefont {van Heel}}, \ and\
  \bibinfo {author} {\bibfnamefont {J.}~\bibnamefont {Kohler}},\ }\href@noop {}
  {\bibfield  {journal} {\bibinfo  {journal} {Phys. Rev. Lett.}\ }\textbf
  {\bibinfo {volume} {94}},\ \bibinfo {pages} {195501} (\bibinfo {year}
  {2005})}\BibitemShut {NoStop}%
\bibitem [{\citenamefont {Wolters}\ \emph {et~al.}(2013)\citenamefont
  {Wolters}, \citenamefont {Sadzak}, \citenamefont {Schell}, \citenamefont
  {Schroder},\ and\ \citenamefont {Benson}}]{BensonPRL2013}%
  \BibitemOpen
  \bibfield  {author} {\bibinfo {author} {\bibfnamefont {J.}~\bibnamefont
  {Wolters}}, \bibinfo {author} {\bibfnamefont {N.}~\bibnamefont {Sadzak}},
  \bibinfo {author} {\bibfnamefont {A.~W.}\ \bibnamefont {Schell}}, \bibinfo
  {author} {\bibfnamefont {T.}~\bibnamefont {Schroder}}, \ and\ \bibinfo
  {author} {\bibfnamefont {O.}~\bibnamefont {Benson}},\ }\href@noop {}
  {\bibfield  {journal} {\bibinfo  {journal} {Phys. Rev. Lett.}\ }\textbf
  {\bibinfo {volume} {110}},\ \bibinfo {pages} {027401} (\bibinfo {year}
  {2013})}\BibitemShut {NoStop}%
\bibitem [{\citenamefont {Eichhammer}\ \emph {et~al.}(2015)\citenamefont
  {Eichhammer}, \citenamefont {Utikal}, \citenamefont {Gotzinger},\ and\
  \citenamefont {Sandoghdar}}]{SandoghdarNJPL2015}%
  \BibitemOpen
  \bibfield  {author} {\bibinfo {author} {\bibfnamefont {E.}~\bibnamefont
  {Eichhammer}}, \bibinfo {author} {\bibfnamefont {T.}~\bibnamefont {Utikal}},
  \bibinfo {author} {\bibfnamefont {S.}~\bibnamefont {Gotzinger}}, \ and\
  \bibinfo {author} {\bibfnamefont {V.}~\bibnamefont {Sandoghdar}},\
  }\href@noop {} {\bibfield  {journal} {\bibinfo  {journal} {New J. Phys.}\
  }\textbf {\bibinfo {volume} {17}},\ \bibinfo {pages} {083018} (\bibinfo
  {year} {2015})}\BibitemShut {NoStop}%
\bibitem [{\citenamefont {Frantsuzov}\ \emph {et~al.}(2009)\citenamefont
  {Frantsuzov}, \citenamefont {Volk\'an-Kacs\'o},\ and\ \citenamefont
  {Jank\'o}}]{FrantsuzovPRL2009}%
  \BibitemOpen
  \bibfield  {author} {\bibinfo {author} {\bibfnamefont {P.~A.}\ \bibnamefont
  {Frantsuzov}}, \bibinfo {author} {\bibfnamefont {S.}~\bibnamefont
  {Volk\'an-Kacs\'o}}, \ and\ \bibinfo {author} {\bibfnamefont
  {B.}~\bibnamefont {Jank\'o}},\ }\href@noop {} {\bibfield  {journal} {\bibinfo
   {journal} {Phys. Rev. Lett.}\ }\textbf {\bibinfo {volume} {103}},\ \bibinfo
  {pages} {207402} (\bibinfo {year} {2009})}\BibitemShut {NoStop}%
\bibitem [{\citenamefont {Frantsuzov}\ \emph {et~al.}(2013)\citenamefont
  {Frantsuzov}, \citenamefont {Volk\'an-Kacs\'o},\ and\ \citenamefont
  {Jank\'o}}]{FrantsuzovNL2013}%
  \BibitemOpen
  \bibfield  {author} {\bibinfo {author} {\bibfnamefont {P.~A.}\ \bibnamefont
  {Frantsuzov}}, \bibinfo {author} {\bibfnamefont {S.}~\bibnamefont
  {Volk\'an-Kacs\'o}}, \ and\ \bibinfo {author} {\bibfnamefont
  {B.}~\bibnamefont {Jank\'o}},\ }\href@noop {} {\bibfield  {journal} {\bibinfo
   {journal} {Nano Lett.}\ }\textbf {\bibinfo {volume} {13}},\ \bibinfo {pages}
  {402} (\bibinfo {year} {2013})}\BibitemShut {NoStop}%
\bibitem [{\citenamefont {Volk\'an-Kacs\'o}\ \emph {et~al.}(2010)\citenamefont
  {Volk\'an-Kacs\'o}, \citenamefont {Frantsuzov},\ and\ \citenamefont
  {Jank\'o}}]{VolkanNL2010}%
  \BibitemOpen
  \bibfield  {author} {\bibinfo {author} {\bibfnamefont {S.}~\bibnamefont
  {Volk\'an-Kacs\'o}}, \bibinfo {author} {\bibfnamefont {P.~A.}\ \bibnamefont
  {Frantsuzov}}, \ and\ \bibinfo {author} {\bibfnamefont {B.}~\bibnamefont
  {Jank\'o}},\ }\href@noop {} {\bibfield  {journal} {\bibinfo  {journal} {Nano
  Lett.}\ }\textbf {\bibinfo {volume} {10}},\ \bibinfo {pages} {2761} (\bibinfo
  {year} {2010})}\BibitemShut {NoStop}%
\bibitem [{\citenamefont {Busov}\ and\ \citenamefont
  {Frantsuzov}(2019)}]{BusovOS2019}%
  \BibitemOpen
  \bibfield  {author} {\bibinfo {author} {\bibfnamefont {V.~K.}\ \bibnamefont
  {Busov}}\ and\ \bibinfo {author} {\bibfnamefont {P.~A.}\ \bibnamefont
  {Frantsuzov}},\ }\href@noop {} {\bibfield  {journal} {\bibinfo  {journal}
  {Opt. Spectr.}\ }\textbf {\bibinfo {volume} {126}},\ \bibinfo {pages} {70 }
  (\bibinfo {year} {2019})}\BibitemShut {NoStop}%
\end{thebibliography}%

\end{document}